\documentclass[runningheads]{llncs}
\usepackage[T1]{fontenc}
\usepackage{graphicx}
\usepackage{multirow}
\usepackage{array}

\begin{document}
\title{Auxiliary Input in Training: Incorporating Catheter Features into Deep Learning Models for ECG-Free Dynamic Coronary Roadmapping}
\titlerunning{Auxiliary Input in Training}

\author{Yikang Liu\inst{1}\orcidID{0000-0003-1069-1215} \and
Lin Zhao\inst{1} \and
Eric Z. Chen\inst{1} \and
Xiao Chen\inst{1} \and
Terrence Chen\inst{1} \and
Shanhui Sun\inst{1}
}

\authorrunning{Y. Liu et al.}

\institute{United Imaging Intelligence, Boston, MA, USA\\
\email{shanhui.sun@uii-ai.com}}
%
%
\maketitle              
\begin{abstract}
Dynamic coronary roadmapping is a technology that overlays the vessel maps (the "roadmap") extracted from an offline image sequence of X-ray angiography onto a live stream of X-ray fluoroscopy in real-time. It aims to offer navigational guidance for interventional surgeries without the need for repeated contrast agent injections, thereby reducing the risks associated with radiation exposure and kidney failure. The precision of the roadmaps is contingent upon the accurate alignment of angiographic and fluoroscopic images based on their cardiac phases, as well as precise catheter tip tracking. The former ensures the selection of a roadmap that closely matches the vessel shape in the current frame, while the latter uses catheter tips as reference points to adjust for translational motion between the roadmap and the present vessel tree. Training deep learning models for both tasks is challenging and underexplored. However, incorporating catheter features into the models could offer substantial benefits, given humans heavily rely on catheters to complete the tasks. To this end, we introduce a simple but effective method, auxiliary input in training (AIT), and demonstrate that it enhances model performance across both tasks, outperforming baseline methods in knowledge incorporation and transfer learning. 

\end{abstract}
%
%
%
\section{Introduction}
\begin{figure}
    \centering
    \includegraphics[width=0.8\textwidth]{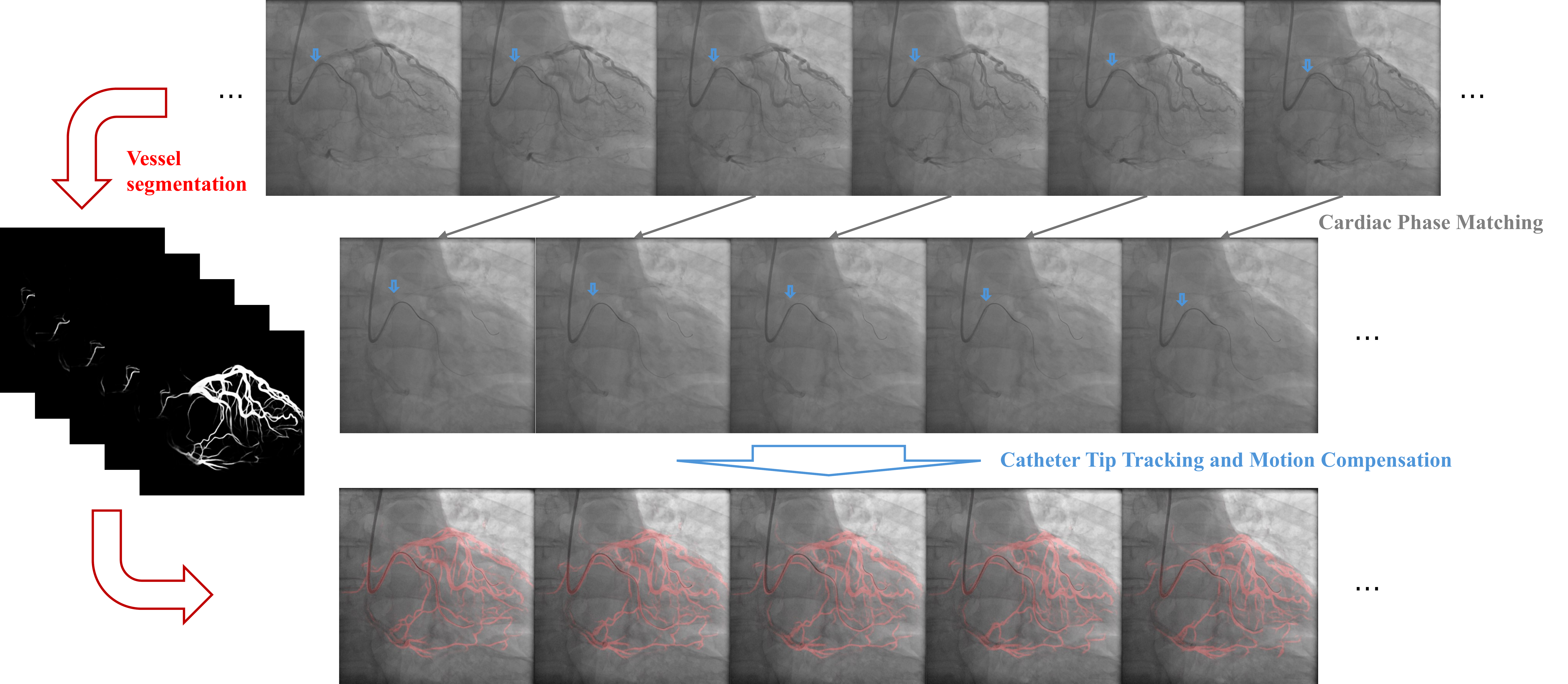}
    \caption{The Workflow of Dynamic Coronary Roadmapping. }
    \label{fig:roadmapping}
\end{figure}

X-ray angiographic image sequences are frequently used in interventional cardiology to assist with the navigation of devices in coronary arteries during procedures such as angioplasty or stent placement. However, it exposes patients to certain risks, including radiation exposure and potential kidney failure due to the use of contrast agents \cite{piayda2018dynamic}. Dynamic coronary roadmapping (Fig. \ref{fig:roadmapping}) is a technology aimed at minimizing the doses of radiation and contrast agent required. It works by superimposing a live contrast-free X-ray fluoroscopic image of the patient with a detailed coronary artery map (the "roadmap"), which is obtained in advance through X-ray angiography. The roadmap is updated real-time accounting for the movement of the heart and the patient's breathing. This approach effectively reduces the necessity for repeated angiography \cite{piayda2018dynamic}.

Dynamic coronary roadmapping can be implemented with three modules (Fig. \ref{fig:roadmapping}): vessel segmentation, cardiac phase matching, and catheter tip detection and tracking. The vessel segmentation module extracts vessel masks from angiographic videos. As coronary vessels deform in cycles due to heartbeats, the cardiac phase matching module ensures the similarity of vessel shapes in the roadmap and live image by selecting the angiographic video frame that best matches the cardiac phase of the live fluoroscopic image. 
The catheter tip detection and tracking module locates the tip of the guiding catheter, which remains stationary in the vessel and serves as a reference for heart and breathing motion, in both the angiographic and fluoroscopic videos. Integrating these components, the system overlays the vessel mask from the selected angiographic frame onto the live image after compensating for translational motion. This overlay serves as the dynamic navigational roadmap.

This paper focuses on developing deep learning models for the cardiac phase matching and catheter tip tracking functions. Humans naturally leverage information from the catheter body to assist with both cardiac phase matching and catheter tip tracking, due to their challenging natures. In cardiac phase matching, the invisibility of vessels in fluoroscopy images makes it difficult to identify the cardiac phase. However, in both angiographic and fluoroscopic image sequences, catheter moves and deforms periodically with heartbeat. A person can compare catheter position and shape across two videos and match their cardiac phases. In catheter tip tracking, the challenges arise from the tip is frequently obscured by contrast agents or other devices. In this case, an individual can approximate their location by examining the shapes of the catheter in adjacent frames where the tips are visible. However, leveraging catheter information proves challenging for deep learning models. As we will demonstrate in this paper, models struggle to converge or learn shortcut rules \cite{hermann2023foundations,geirhos2020shortcut} that fail to generalize in more challenging scenarios (e.g. the presence of thick contrast agents). 

Training deep learning models to leverage catheter body features can be seen as an application of domain knowledge incorporation \cite{xie2021survey,dash2022review} or transfer learning \cite{zhuang2020comprehensive}. Typical approaches for medical images under this umbrella include fine-tuning a pretrained network \cite{8466816,shin2016deep}, multi-task learning \cite{zhao2020tripartite,bakalo2019classification}, custom neural network architecture \cite{guan2018diagnose,wang2018tienet}, teacher-student models \cite{han2020deep,hu2022teacher}, generative style transfer \cite{zhao2020tripartite}, and constraining image features with attention maps \cite{li2019attention,cui2020collaborative}. 

In the context of catheter tip tracking with deep learning, prior research has enhanced model capabilities with a dedicated branch for predicting the catheter mask \cite{demoustier2023contrack,ma2020dynamic} and custom architectures with emphasis on catheter motion \cite{demoustier2023contrack}. In contrast, incorporating catheter knowledge into cardiac phase matching models is less explored. Prior research has either used ECG signals to match cardiac phases \cite{ma2020dynamic} or analyzed the catheter curvature to assess the similarity between two frames' cardiac phases \cite{piayda2018dynamic}. These methods face limitations, as ECG signals may not always be accessible in clinical settings, and manually engineered features tend to be less reliable than machine learning approaches, especially when the catheter's shape is complicated by foreshortening or overlaps with other objects. Ciusdel et al. \cite{ciusdel2020deep} developed a deep learning model to identify cardiac phase, but its application is limited to angiographic images with visible vessels. 

In this paper, we introduce Auxiliary Input in Training (AIT), a simple yet effective method that leverages catheter masks as auxiliary signals to incorporate catheter features into deep learning models for cardiac phase matching and catheter tip tracking, thereby enhancing representation quality and accelerating convergence. 
By appending the catheter mask as an additional input channel and gradually ablate it to a zero matrix during training, AIT facilitates the integration of catheter information. This approach has enabled us to create, to our knowledge, the first end-to-end deep learning framework capable of accurately matching coronary X-ray frames by cardiac phases. 
Moreover, we demonstrate that AIT also improves catheter tip tracking models, outperforming baseline methods in knowledge incorporation or transfer learning despite its simplicity. 

\section{Methodology}
In this section, we first introduce the concept of AIT, then formally define the problems of cardiac phase matching and catheter tip tracking, and present our models and loss functions. More details are in the supplementary material.
\subsubsection{AIT}
Suppose we want to train a deep learning model $f: x \mapsto y$ on a dataset $D=\{x,y\}$, but the model is hard to train due to the unsmoothness of the loss landscape or a complicated relationship between $x$ and $y$. As a consequence, directly training $f$ on $D$ may take very long to converge or converge to minima that fail to generalize to the test dataset (e.g. shortcut learning \cite{hermann2023foundations,geirhos2020shortcut}). The key idea of AIT is to introduce an auxiliary input $z$ to guide the training process so that the training converges faster and better representations are learned. More specifically, we initially train $f(x;z)$ on a dataset $D_z = \{x,y,z\}$, and gradually ablate $z$ throughout the training process so that the model's reliance from the auxiliary information $z$ is transferred back to the primary input $x$, which is the only variable needed for inference. The auxiliary input $z$, which though can be inferred from $x$, has a more obvious relationship to $y$ (with respect to the network's architecture or inductive bias) or is an indispensable step to infer $y$ with $x$ based on prior knowledge (e.g. humans rely on catheter shape consistency to identify cardiac phases and track tips). 

In the applications of cardiac phase matching and catheter tip tracking, $z$ is a binary catheter mask, which is concatenated with the input image $x$ along the channel dimension. The ablation of $z$ is done by adding Gaussian noise and concurrently decreasing the signal magnitude:
\[ \tilde{z} = (1-\mathrm{\alpha})((1-\mathrm{\alpha})z + \mathrm{\alpha}\mathcal{N}(0, 1)) \]
, where $\tilde{z}$ is the ablated $z$ and $\alpha$ is a parameter that adjusts the intensity of the ablation. Throughout the training process, $\alpha$ is progressively increased from 0 to 1. At the point where $\alpha = 1$, $z$ is entirely ablated into a matrix of zeros, thus becoming unnecessary for inference. In our default setting, we increment $\alpha$ by 0.1 at each step. After achieving network convergence at a given ablation level without any signs of overfitting, we escalate to the subsequent ablation level.

\subsubsection{Cardiac Phase Matching}\label{cpm}
Given a sequence of recorded cardiac angiographic images $\{I_i^A\}$ and a live fluoroscopic image stream $\{I_i^F\}$, the cardiac phase matching function finds the image in $\{I_i^A\}$ that best matches the cardiac phase of the current image $I_i^F$ in real-time. We achieve this by using a CNN encoder to extract features from each image and a temporal neural network to infer temporal relations between image features, for which we experimented with both ConvLSTM and Transformer backbones to show the effectiveness of AIT on different architectures. The model outputs a feature vector $v_i$ for each image and the cosine similarity between two feature vectors measures how close the cardiac phases of the corresponding images are. 

The CNN-ConvLSTM model (hereafter denoted as CNN-C) comprises a series of alternating UNetResBlock \cite{cardoso2022monai} and ConvLSTM \cite{shi2015convolutional}
layers, followed by a global max pooling and a fully connected layer to transform a 4D image tensor into a 1D feature vector. $\{I_i^A\}$ and $\{I_i^F\}$ are concatenated along the temporal dimension and sequentially fed into the model (Fig. S1). 

The CNN-Transformer model (hereafter denoted as CNN-T) comprises a ResNet encoder and stacked attention layers. The outputs from the ResNet are flattened into 1D vectors before being passed to the attention layers. The attention layers run with self-attention for $\{I_i^A\}$. For real-time inference of $\{I_i^F\}$, the features extracted from the current fluoroscopic image are used as the query vector, while features from previous frames are used as key and value vectors. 

Both models were trained with a triplet loss
\[\mathcal{L} = \max(-S(v^F, v_p^A) + S(v^F, v_n^A) + \epsilon, 0)\]
$S$ denotes the cosine similarity. $v^F$ is the feature vector of a fluoroscopic image $I_i^F$. $v_p^A$ is the feature vector of an angiographic image image $I_i^A$ with the same cardiac phase, whereas $v_n^A$ is the feature vector of one with a different cardiac phase. $\epsilon$ represents a positive margin, which we set to 0.8 in all experiments.

\subsubsection{Catheter Tip Tracking}
Catheter tip tracking involves determining the coordinates $(x, y)$ of a catheter tip within an image, based on one or multiple previous images and their corresponding tip coordinates. Similar to cardiac phase matching, we used both CNN-C and CNN-T models to show the effectiveness of AIT on different architectures.

The CNN-C model is a UNet with ConvLSTM layers in the skip connections. 
The input is a sequence of 3-channel tensors, with each channel containing the reference image (the image where tip location is known), the reference tip heatmap, and the current image to inference. The tensors are sequentially inputted into the network, which then outputs a tip heatmap for each frame. 

The CNN-T model is similar to STARK-S\cite{yan2021learning}. It takes a template obtained by cropping the reference image around the tip and a search image, and passes them through a ResNet encoder. The encoder's outputs are flattened, concatenated, and then forwarded to a transformer encoder. Subsequently, a trainable target query and the transformer encoder's output are sent to a transformer decoder. The resulting output is further processed by a CNN head for heatmap regression. Similar to \cite{demoustier2023contrack}, three templates are used to adapt to variations in the tip's appearance, which includes the initial template and two from the latest tracked tips, selected if their probabilities in the heatmaps exceeding a certain threshold (0.5). In AIT, catheter masks are concatenated with both the templates and search images.

Both models were trained with the L1 loss between the predicted heatmaps and the labels. 

\section{Experiments}

\subsubsection{Baseline Methods}
In addition to vanilla supervised learning, we also compared AIT with three other methods on both the cardiac phase matching and the catheter tip tracking tasks (hereafter denoted as CPM and CTT, respectively).

The first method (denoted as FT) fine-tunes a model trained for catheter segmentation. In the CPM task, the segmentation network builds upon the original CNN-C or CNN-T architecture (Section \ref{cpm}) and appends a CNN decoder after the last ConvLSTM/Attention layer for mask regression. In the CTT task, the segmentation models share the same structure as the tracking models. 

The second method employs a multi-task learning (MTL) approach, wherein the models feature two branches simultaneously trained to predict both catheter masks and task-specific outputs. In the CPM task, the catheter segmentation branch uses the same CNN decoder structure in the FT method. In the CTT task, the catheter segmentation branch parallels the UNet decoder (in the CNN-C) or the heatmap regression head (in the CNN-T), having the same structures. 

The third method uses a teacher-student (T-S) model, where the teacher network, pre-trained for catheter segmentation, guides the student (target) model with the maximum mean discrepancy (MMD) loss on the student's and teacher's features. The teacher models share the same architectures with the segmentation networks in FT. For fair comparison with MTL, the MMD loss is applied to the features before the segmentation branch, where networks start to use separate features for segmentation and target tasks.

In all experiments, we set the learning rates to $10^{-5}$ and used the Adam optimizer, with betas configured to 0.9 and 0.999. 

\subsubsection{Ablation Studies}
We investigated how AIT was affected by the percentage of data that trained with auxiliary inputs. This question is important since acquiring extra labels can be expensive. We run AIT on both tasks with partial inclusions (20\%, 40\%, 60\%, and 80\%) of catheter masks, where zero matrices were used as placeholders for the missing catheter masks.

\subsubsection{Datasets and Evaluation Metrics} 
All experiments were conducted using in-house data, obtained with institutional committee approval. The datasets for CPM and CTT contain 2483 pairs of angiographic and fluoroscopic videos (174228 frames in total) and 4098 videos (255432 frames) respectively, with frame rates equal to 7.5, 15, or 30 fps and image sizes range from $492 \times 492$ to $624 \times 624$ after normalizing pixel spacing to isotropic 0.2 mm. Cardiac phases, catheter tips, and catheter masks were manually labeled. The datasets were divided into training, validation, and testing sets with a 7:2:1 ratio. 

The performance of models on the CPM task is assessed with matching accuracy, defined as the temporal distance between the predicted frame and the nearest ground-truth frame (multiple frames may exhibit the same cardiac phase due to the periodic nature of heartbeats). For the distance metric, we employed two units of measurement: frame counts and the percentage of a cardiac cycle. For example, if the distance is 2 frames within a cardiac cycle spanning 12 frames, the corresponding percentage is calculated as 1/6 or 16.7\%. To evaluate the performance on the CTT task, we deemed a tracking attempt successful if the distance between prediction and ground-truth did not exceed 2 mm. This threshold is consistent with the outside diameter of a guiding catheter \cite{ojha2023catheters} and meets the conventional requirements of roadmap accuracy \cite{faranesh2013integration,piayda2018dynamic}. Using this criterion, we calculated the precision (P) and recall (R) for tracking. We also calculated the distance mean and standard deviation for true positives (TPs) and all cases.


\section{Results and Discussion}
Both CNN-C and CNN-T backbones have inference times (on an Nvidia V100 GPU) under 25 ms and 40 ms for the CPM and CTT tasks, respectively, satisfying the requirements for clinical application (15 fps).

\begin{table}
\centering
\caption{Performance on the Cardiac Phase Matching Task. $\alpha$ is the ablation strength on auxiliary inputs. \textbf{Bold} font indicates the best method (paired t-tests, $p<0.05$) excluding intermediate AIT results ($\alpha \neq 1$). }\label{tab:CPM}

\begin{tabular}{lcc|cc}
\hline
\multirow{2}{*}{Methods} & \multicolumn{2}{c}{CNN-ConvLSTM} & \multicolumn{2}{c}{CNN-Transformer} \\
\cline{2-5}
 & dist(frame)$\downarrow$ & dist(\%)$\downarrow$ & dist(frame)$\downarrow$ & dist(\%)$\downarrow$ \\
\hline
Vanilla &3.26±1.87	&25.00±14.42	&3.25±1.88	&25.00±14.45 \\
FT &3.19±1.85	&24.01±13.81	&3.21±1.84	&24.49±14.13\\
MTL &3.25±1.87	&24.99±14.43	&3.30±1.91	&24.46±14.10\\
T-S &2.35±1.35	&17.42±10.08	&2.44±1.41	&19.05±10.98\\
\hline
AIT ($\alpha=0$) &0.92±0.54	&7.26±4.22	&0.87±0.51	&6.85±4.00 \\
AIT ($\alpha=0.5$) &0.96±0.57	&7.69±4.45	&0.97±0.59	&7.75±4.59 \\
AIT ($\alpha=0.8$) &0.96±0.56	&7.34±4.34	&0.95±0.54	&7.32±4.31 \\
AIT (final) &\textbf{0.89±0.51}	&\textbf{7.01±4.08}	&\textbf{0.85±0.51}	&\textbf{6.72±3.98} \\
\hline
\end{tabular}

\end{table}

\begin{figure}[!ht]
    \centering
    \includegraphics[width=0.8\textwidth]{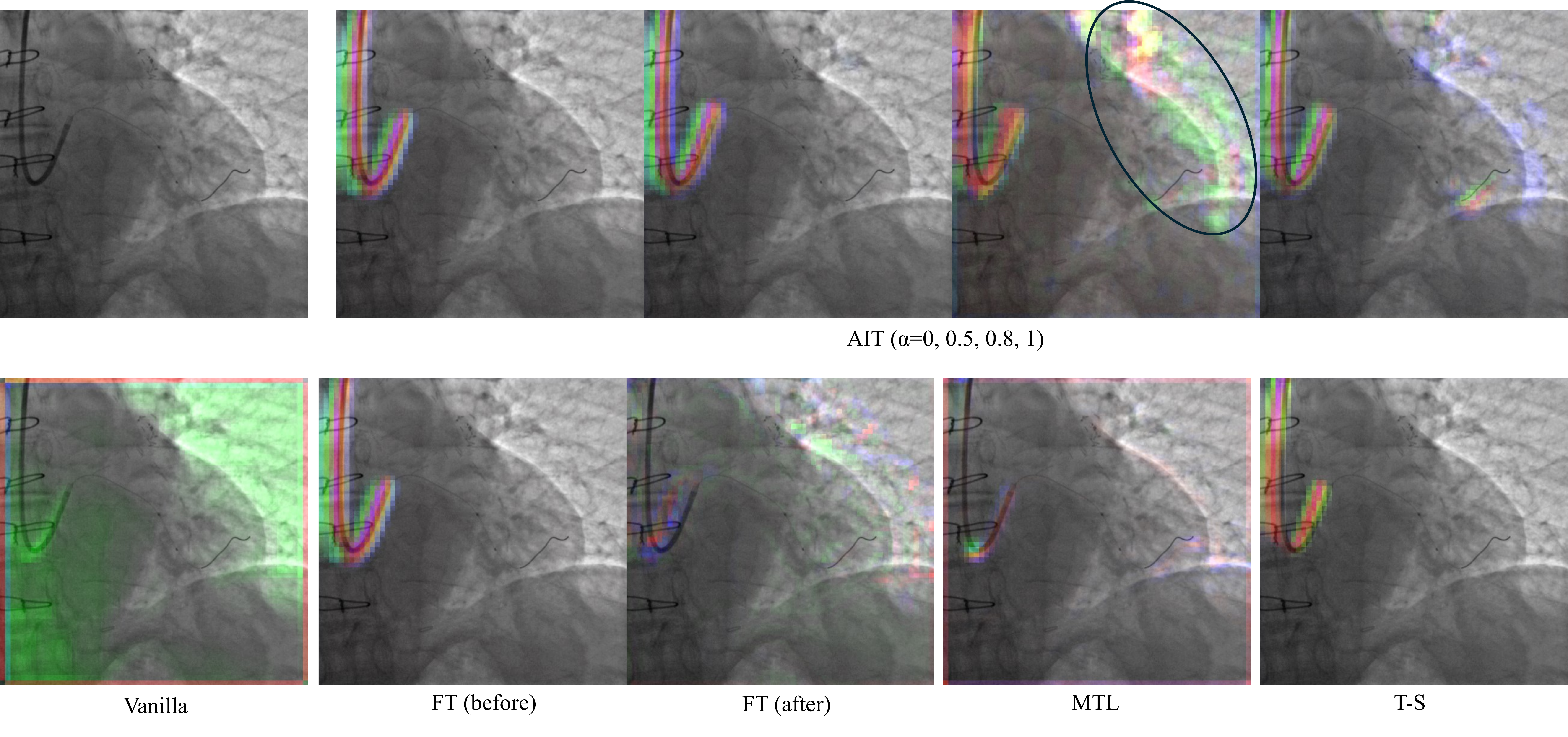}
    \caption{Visualization of Features Learned by Different Methods in the CPM task. First row (left to right): original image and AIT ($\alpha$=0, 0.5, 0.8, 1). Second row (left to right): vanilla supervised learning, FT (before and after fine-tuning), MTL, and T-S.}
    \label{fig:pm-vis}
\end{figure}

The performance of all the methods on the CPM task, along with AIT performance at different ablation strengths, is shown in Table \ref{tab:CPM}. We trained CNN-C and CNN-T models using AIT and other methods for 300 and 800 epochs, respectively. It was observed that both networks failed to converge under the vanilla, FT, and MTL strategies, as evidenced by a distance metric around 25\%, indicating that the networks were essentially making random guesses. The T-S method produced predictions above random chance, yet its performance was significantly inferior to that of AIT (paired t-test, $p<0.05$) and did not meet clinical standards. To better understand the underlying causes, we visualized the features after the second ConvLSTM block in the CNN-C backbone (Fig. \ref{fig:pm-vis}), which was done by concatenating the magnitude of the first three principal components as RGB channels. It can be observed that AIT($\alpha=1$) was able to learn strong features related to cardiac phase, located at catheter, wire, heart contour (circled in Fig. \ref{fig:pm-vis}), whereas other methods learned much weaker features. These observations indicate that AIT is able to facilitate model convergence by incorporating catheter features.

Additionally, it is observed that AIT's performance dip initially with the start of auxiliary input ablation but improved near the end of the ablation schedule. Together with the feature maps in Fig. \ref{fig:pm-vis}, it suggests that the model initially relied on catheter features, which became weaker due to mask ablation. Subsequently, the model adapted by leveraging alternative features to offset the weakened catheter features, resulting in more robust predictions than those based solely on catheter information.

AIT achieved sub-frame average accuracy with both networks, indicating its potential to replace ECG-based phase matching methods. However, to assess its practical clinical use compared to the ECG-based method, the accuracy needs further evaluation by measuring the distance between the overlaid vessels and the interventional devices (e.g. guide wires) within the vessels, as demonstrated in \cite{ma2020dynamic}. This practical assessment will be included in other future work.

\begin{table}[th!]
\centering
\caption{Performance on the Catheter Tip Tracking Task. $\alpha$ is the ablation strength on auxiliary inputs. \textbf{Bold} font indicates the best method (paired t-tests, $p<0.05$) excluding intermediate AIT results ($\alpha \neq 1$). }\label{tab:CCT}
\begin{tabular}{lcccc|cccc}
\hline
\multirow{2}{*}{Methods} & \multicolumn{4}{c}{CNN-ConvLSTM} & \multicolumn{4}{c}{CNN-Transformer} \\
\cline{2-9}
 & P(\%)$\uparrow$ & R(\%)$\uparrow$ & dist(TP)$\downarrow$ & dist(all)$\downarrow$ & P(\%)$\uparrow$ & R(\%)$\uparrow$ & dist(TP)$\downarrow$ & dist(all)$\downarrow$\\
\hline
Vanilla &91.3	&94.5 &0.90±0.63 & 1.67±2.88 & 92.5	&95.2 &0.92±0.62 & 1.45±2.24\\
FT &92.6	&95.9 &0.89±0.66 & 1.41±2.16 &93.0	&96.1 &0.91±0.57 & 1.33±1.81\\
MTL &90.5	&94.0 &0.91±0.79 & 1.75±2.98 &91.6	&94.1 &0.93±0.65 & 1.56±2.41\\
T-S &94.7	&97.4 &0.89±0.52 & 1.26±1.87 &95.5	&\textbf{98.0} &0.91±0.57 & 1.13±1.36\\
\hline
AIT ($\alpha=0$)  &99.7	&99.8 &0.87±0.49 & 0.89±0.62 &99.7	&99.8 &0.90±0.51 & 0.91±0.61\\
AIT ($\alpha=0.5$)  &98.2	&99.0  &0.88±0.53 & 0.96±0.88 &97.8	&98.7 &0.91±0.54 & 1.04±1.08\\
AIT ($\alpha=0.8$)  &97.7	&98.6  &0.88±0.51 & 1.00±1.04 &97.2	&98.5 &0.91±0.53 & 1.08±1.22\\
AIT (final)  &\textbf{96.9}	&\textbf{97.6}  &\textbf{0.88±0.50} & \textbf{1.10±1.48} &\textbf{97.1}	&97.8 &\textbf{0.90±0.54} & \textbf{1.08±1.23}\\
\hline
\end{tabular}
\end{table}

In the CTT task, AIT achieved the highest performance using the CNN-C backbone and ranked as either the best or second-best using the CNN-T backbone (Table \ref{tab:CCT}). Generally, MTL negatively impacted performance (consistent with \cite{demoustier2023contrack}), while other approaches contributed positively. AIT performance decreased with the progression of auxiliary input ablation, due to ease of inferring catheter tip positions from clean catheter masks. However, it still significantly outperformed the vanilla supervised learning method (paired t-tests, $p<0.05$), indicating that catheter features were effectively incorporated into the models even in the absence of catheter masks as inputs, thereby improving performance. Furthermore, it should be noted that vanilla CNN-T resembles ConTrack without multitask, flow, or multi-templates (refer to Table 2 in \cite{demoustier2023contrack}).
AIT improved the average tracking distance by 25.5\% to 1.08 mm, compared to ConTrack's improvement by 24.9\% to 1.63 mm (Table 2 in \cite{demoustier2023contrack}). Although these models were trained and tested on different datasets, the results suggest that AIT's approach of integrating catheter body information might achieve performance levels similar to those of specifically designed neural networks.

\begin{figure}
    \centering
    \includegraphics[width=0.9\textwidth]{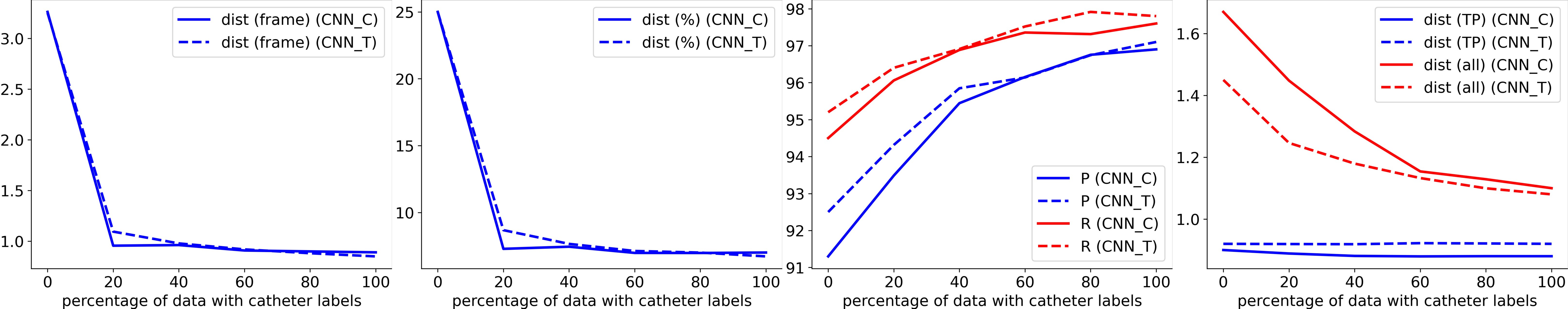}
    \caption{Ablation Study}
    \label{fig:ablation}
\end{figure}

Finally, we demonstrate in the ablation study that even a small percentage of auxiliary inputs can yield significant benefits, especially in the CPM task, as shown in Fig. \ref{fig:ablation}. 

\section{Conclusion}
This study introduces a straightforward yet effective approach, Auxiliary Input in Training (AIT), for incorporating prior knowledge into deep learning models. We applied this method to train models for cardiac phase matching and catheter tip tracking—two demanding tasks in dynamic coronary roadmapping—and showcased its efficacy in enhancing performance across both tasks. Despite its simplicity, AIT's superior performance stands out in comparison to other techniques, underscoring its value in complex medical imaging tasks.

\begin{credits}

\subsubsection{\discintname}
The authors have no competing interests to declare that are
relevant to the content of this article.
\end{credits}

\clearpage
%





\appendix
\section*{Supplementary Material}

\newcolumntype{L}[1]{>{\raggedright\arraybackslash}p{#1\textwidth}}
\newcolumntype{C}[1]{>{\centering\arraybackslash}p{#1\textwidth}}
\newcolumntype{T}[1]{>{\raggedright\arraybackslash\vtop{\vskip0pt plus 1filll\hbox{##1}}}}

\renewcommand{\thefigure}{S\arabic{figure}}
\renewcommand{\thetable}{S\arabic{table}}

\setcounter{figure}{0}
\setcounter{table}{0}

\section{Network Architectures}
\subsection{CNN-ConvLSTM}

\begin{figure}
    \centering
    \includegraphics[width=\textwidth]{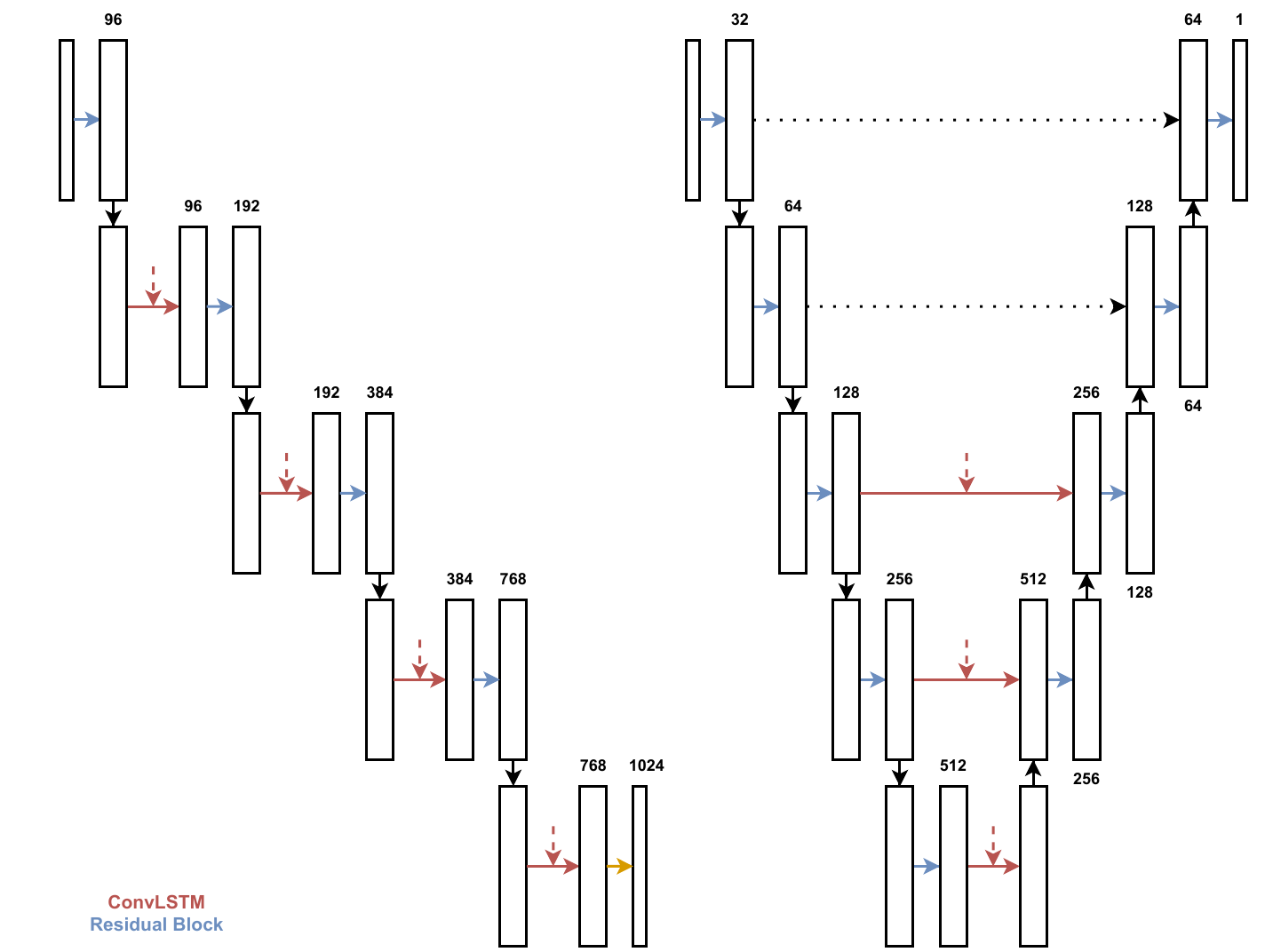}
    \caption{\textbf{Architectures of CNN-ConvLSTM Networks.} The left is the model for cardiac phase matching, and the right is for catheter tip tracking. Numbers above tensors are channel numbers.}
    \label{fig:model}
\end{figure}

Fig. \ref{fig:model} shows the model architectures of CNN-ConvLSTM networks for cardiac phase matching (CPM) and catheter tip tracking (CTT). 

The CPM model comprises a series of alternating UNetResBlock \cite{cardoso2022monai}\footnote{https://docs.monai.io/en/stable/networks.html, v1.2.0} layers, ConvLSTM \cite{shi2015convolutional}\footnote{https://github.com/ndrplz/ConvLSTM\_pytorch} 
layers, and downsampling layers (implemented by 1x1 convolution layers with stride=2), followed by a global max pooling and a fully connected layer to transform a 4D image tensor into a 1D feature vector. A sequence of recorded cardiac angiographic images and a live fluoroscopic image stream are concatenated along the temporal dimension and sequentially fed into the model. 

The CTT model is a UNet with ConvLSTM layers in the skip connections. 
The input is a sequence of 3-channel tensors, with each channel containing the reference image (the image where tip location is known), the reference tip heatmap, and the current image to inference. The tensors are sequentially inputted into the network, which then outputs a tip heatmap for each frame.

The input and output channel numbers of UNetResBlocks and ConvLSTM layers are shown in Fig. \ref{fig:model}. Other hyperparameters are shown in Table \ref{tab:param}. 

\begin{table}
\centering
\caption{Hyperparameters of UNetResBlock and ConvLSTM Layers. }\label{tab:param}
\begin{tabular}{L{0.25}|C{0.25}|L{0.25}|C{0.25}}
\hline
\multicolumn{2}{c}{UNetResBlock} & \multicolumn{2}{c}{ConvLSTM} \\
\hline
spatial\_dims &2 & hidden\_dim & output channel\#\\
kernel\_size &3	& kernel\_size & 3 \\
norm\_name & None &  bias& True \\
act\_name  & relu   &   layers  & \begin{tabular}{c} 3 (the last) \\ 1 (others) \end{tabular}\\
\hline
\end{tabular}
\end{table}

\subsection{CNN-Transformer}
The CNN-Transformer model for cardiac phase matching comprises a ResNet encoder \cite{he2016deep} (the part of ResNet-50 before the average pooling layer) and stacked attention layers. The outputs from the last two stages of the ResNet encoder are averaged globally and flattened and concatenated into 1D vectors (with a dimension of 3072) before being passed to the attention layers. Attention layers were implemented with \textit{torch.nn.MultiheadAttention}\footnote{v2.0}, with \textit{embed\_dim}=3072, \textit{num\_heads}=4, and other parameters were set to defaults. Five attention layers were used with residual connections. The attention layers run with self-attention for recorded cardiac angiographic images. For real-time inference of the live fluoroscopic image stream, the features extracted from the current fluoroscopic image are used as the query vector, while features from previous frames are used as key and value vectors. 

The CNN-Transformer model for cardiac tip tracking has the same backbone as STARK-S50\cite{yan2021learning}, except for that one heatmap was generated indicating the tip location instead of two heatmaps for corners of the bounding box. The template size is 64x64. 
\clearpage

\bibliographystyle{splncs04}
\bibliography{Paper-1135}

\end{document}